# Thickness-dependent Local Surface Electronic Structures of Homoepitaxial SrTiO$_3$ Thin Films

Takeo Ohsawa,[1*] Katsuya Iwaya,[1] Ryota Shimizu,[1,2] Tomihiro Hashizume,[1,3,4] and Taro Hitosugi[1]

[1] WPI-Advanced Institute for Materials Research, Tohoku University, Sendai 980-8577, Japan.

[2] Department of Chemistry, The University of Tokyo, Bunkyo, Tokyo 113-0033, Japan.

[3] Advanced Research Laboratory, Hitachi, Ltd., Hatoyama, Saitama 350-0395, Japan.

[4] Department of Physics, Tokyo Institute of Technology, Meguro, Tokyo 152-8551, Japan.

*ohsawa@wpi-aimr.tohoku.ac.jp

We have investigated the atomically-resolved substrate and homoepitaxial thin film surfaces of SrTiO$_3$(001) using low-temperature scanning tunneling microscopy/spectroscopy (STM/STS) combined with pulsed laser deposition (PLD). It was found that annealing at 1000 °C in an oxygen partial pressure of 1×10$^{-6}$ Torr, which is a typical annealing treatment for the preparation of SrTiO$_3$ substrates, unexpectedly resulted in a disordered surface on an atomic scale. In contrast, homoepitaxial SrTiO$_3$ thin films grown on this disordered substrate exhibited (2×2) surface reconstructions. The differential conductance spectra, d$I$/d$V$ in STS measurements, revealed a number of surface defects in a 10-unit-cell-thick SrTiO$_3$ film, but much fewer in a 50-unit-cell-thick film. These results indicate that the defect density in the film strongly depends on the film thickness, suggesting non-uniform stoichiometry along the growth direction.

Strontium titanate ($SrTiO_3$) exhibits a wide variety of physical properties, including a metal-insulator transition,[1] superconductivity,[2,3] ferroelectricity,[4] and quantum paraelectricity,[5] which provides a strong incentive to develop functional perovskite-based oxide thin films and heterostructures based on $SrTiO_3$. Recent remarkable advances in heteroepitaxial growth of metal oxides, using pulsed laser deposition (PLD) and molecular beam epitaxy (MBE), have allowed the formation of atomically abrupt interfaces, and such structures have exhibited, for example, the presence of a high-mobility quasi two-dimensional electron gas (q-2DEG),[6,7,8,9] magnetism,[10] and superconductivity.[11,12] As the interfaces of such heterostructures are of crucial importance, more attention should be paid to the atomic-scale nature of the substrate and film surfaces, in order to gain a better understanding of such phenomena occurring in complex oxide thin films and heterostructures.

One of the most promising techniques for exploring materials on a nanoscale is scanning tunneling microscopy/spectroscopy (STM/STS). To examine the surfaces and growth process of oxide thin films that directly link to the origins of the intriguing properties of epitaxial film surfaces and interfaces, the combination of low-temperature STM/STS with PLD or MBE is an extremely powerful technique. In this context, it is indispensable to investigate clean oxide surfaces, which have not been exposed to air during transfer from the deposition chamber to the STM, to avoid surface contamination that degrades the intrinsic properties of the films. There have been a few room-temperature STM studies on the clean surfaces of oxide thin films such as titanates[13,14] and manganites,[15,16,17] in contrast to a number of the reports on the SrTiO$_3$(001) single-crystal surfaces on which these overlayers are grown.[18,19] To date, however, there exists no report on the low-temperature STM/STS investigation of oxide *thin films* with clean surfaces. Consequently, investigation of the atomic-scale electronic structure of oxide thin films is a vast unexplored field that would contribute to the understanding of the fundamental physics and chemistry of such films.

In this letter, we present a study on the local probing of the surface electronic structures of SrTiO$_3$ thin films, homoepitaxially grown by PLD in a layer-by-layer manner, using low-temperature STM/STS. We first show that a typical epitaxial-growth-ready SrTiO$_3$ substrate with a step-and-terrace surface does not show any atomic order, but rather, exhibits inhomogeneous electronic structures far from the idealized TiO$_2$-terminated SrTiO$_3$(001) surface structure. We then discuss the thickness dependence of the STS results for SrTiO$_3$ thin film surfaces. An epitaxial film with a thickness of 10 unit cells (u.c.) exhibits many randomly distributed surface defects, in contrast to a thick film (50 u.c.) that exhibits rather uniform electronic structures, suggesting the incorporation of defects in the initial stage of thin film growth. This study elucidates the defect distribution and electronic states of oxide thin films with atomic resolution, which plays a fundamental role in the manifestation of various physical properties at the interfaces.

Undoped SrTiO$_3$ thin films were homoepitaxially grown on buffered-HF-etched Nb-doped SrTiO$_3$(001) (Nb: 0.1 at.%, Nb:SrTiO$_3$) substrates[20] using PLD. A KrF excimer laser ($\lambda$ = 248 nm) with a repetition rate of 2 Hz was employed for ablation of single-crystal SrTiO$_3$ as a target material. The laser fluence at the target surface was ~1.2 J/cm$^2$ in our optimized growth conditions (see supporting information). The films were grown in an oxygen pressure of $1\times10^{-6}$ Torr at a temperature of 700 ˚C produced by direct current resistive heating through the samples. Prior to growth, the substrate temperature was held constant at 1000 ˚C for 30 min in an oxygen pressure of $1\times10^{-6}$ Torr, which is a typical treatment used to produce SrTiO$_3$ surfaces with straight step-edges;[9] such epitaxial-growth-ready surfaces are hereafter referred to as epi-ready surfaces.[21] Layer-by-layer SrTiO$_3$ homoepitaxial growth was monitored *in situ* by reflection high-energy electron diffraction (RHEED). After the growth, samples were cooled down to room temperature at a rate of ~3.2 K/sec, followed by immediate transfer to the STM chamber without exposing the sample surface to air. STM/STS measurements were conducted at 78 K with a Pt-Ir tip, and all the STM images were obtained in a constant current mode. Details of our system will be published elsewhere.[22]

We first investigated the epi-ready surface of a Nb:SrTiO$_3$(001) substrate, whose morphology changes with substrate annealing temperature due to the rapid migration of surface atoms, along with the removal of adsorbed surface contaminations.[23] Annealing at 1000 °C in $1\times10^{-6}$ Torr of oxygen, reproducibly provides a straight step-and-terrace surface. Figure 1(a) shows an empty-state STM image of the resulting surface, which clearly indicates equidistant and straight steps consistent with the observed streaky RHEED pattern (inset of Figure 1(a)). The cross-sectional profile, measured normal to the step edge, showed that the step height was 0.4 nm, close to the $SrTiO_3$ lattice constant, as shown in Figure 1(b). The magnified STM image in Figure 1(c), however, indicates the disordered nature of the surface on an atomic scale. This result seems to imply that the typical annealing step conducted prior to oxide film growth allows us to obtain straight step-and-terrace structures,[9] but does *not* produce a structurally-ordered surface on an atomic scale. This suggests that the widely accepted procedure to obtain a perfect epi-ready "$TiO_2$-terminated $SrTiO_3$ surface" for film growth should be reviewed more carefully.

We next grew $SrTiO_3$ thin films on the disordered substrate surface shown in Figure 1(c). Figures 2(a) and (b) show RHEED intensity oscillations monitored during the growth of 10-u.c.- and 50-u.c.-thick films, respectively. Layer-by-layer RHEED oscillations were observed for both films, allowing the number of unit-cell-thick layers to be counted during the deposition. Thus, we stopped the epitaxial growth when the specular spot intensities reached the maxima corresponding to thicknesses of 10 and 50 u.c., respectively.

We show in Figures 2(c)-(f) the STM images of these 10-u.c.- and 50-u.c.-thick films. The wide-area STM images in Figure 2(c) and (d) show step-and-terrace surfaces on which holes and islands with a constant height of 0.4 nm can be clearly observed. The magnified STM images of these two films seen in Figures 2(e) and (f), interestingly, show similar features: a (2×2) surface reconstruction and a number of bright spots embedded in the surface. The (2×2) reconstruction was clearly observed at a relatively low sample-bias-voltage ($V_s$) of ~+2.0 V and can be identified in the fast Fourier transformed (FFT) image (shown in the inset of Figure 2(e)), as well as the RHEED pattern. In addition to the ordered structures, many bright spots possibly associated with defects were visible. It should be noted that the number of these defects in the STM images greatly increased at higher bias voltages in the empty states. Thus, the STM images show a strong $V_s$ dependence, and the (2×2) reconstruction could hardly be seen at ~+3.5 V due to these defects, as shown in Figure 3(a), which is obtained from the same area as in Figure 2(e).

To gain more insight into how the electronic states change as $V_s$ varies, we conducted d$I$/d$V$ measurements, a measure of the local density of states, for both films in the range from 0 to +3.5 V. Figure 3(a) is an STM image of the 10-u.c.-thick film at $V_s$ = +3.5 V where we performed a 128×128 points STS measurement. We can clearly identify the defects in the d$I$/d$V$ map, where bright dots with a typical size of ~1 nm appear, and the number of defects apparently increases with increasing $V_s$ as shown in Figures 3(b) and (c). In addition, the d$I$/d$V$ spectra taken at defect sites clearly show peak structures which can be categorized into three different peaks at around +2.8, +3.0, and +3.2 V in Figure 3(d). In contrast, the peak structures in the d$I$/d$V$ spectra were not seen at the defect-free sites (black curve in Figure 3(d)). We note that these defects are most likely associated with oxygen vacancies because their number substantially increases in a SrTiO$_3$ film grown at a higher growth temperature of 1100 ˚C or decreases in the film grown under a higher oxygen partial pressure. The details will be discussed in separate reports.[24,25]

Similar peak structures in the d$I$/d$V$ spectra were also observed in the epi-ready surface. To compare the defect density on the epi-ready and the 10-u.c.-thick film surfaces, we plot the areal defect density of both surfaces as a function of peak voltage in Figure 3(e). For the epi-ready surface, the areal density was almost uniform for all defects (~1×10$^{13}$ cm$^{-2}$) distributed in a wide voltage range between +2.0 and +3.2 V. In the 10-u.c.-thick film, however, the areal defect density is significantly reduced and the peak voltage is confined to a much narrower range (+2.6 ~ +3.2 V), thus indicating fewer defects and a more uniform spatial defect distribution than for the epi-ready surface.

We then performed the same STS measurements for the 50-u.c.-thick film to investigate the thickness dependence of the surface electronic states. The d$I$/d$V$ maps, surprisingly, revealed the absence of any apparent defects in the range from +2.0 to +3.0 V (Figure 4(a)), which is in clear contrast to the thinner 10-u.c.-thick film shown in Figure 4(b). This result indicates that the thicker film has more uniform electronic states than that of thinner films; in other words, the defect density decreases with increasing film thickness. Indeed, the lattice constant of a thick (100 u.c.) SrTiO$_3$ film grown under the same condition was identical to that of the bulk SrTiO$_3$, supporting the idea that the defect density decreased in the thicker films.

To illustrate the "electronic homogeneity" in a more quantitative way, we present in Figure 4(c) the averaged d$I$/d$V$ spectra with standard deviations that reflect the distribution of peak structures resulting from defects on the surfaces. All the spectra in the STS measurements were averaged for three surfaces. For +2 V < $V_s$ < +3.2 V, the averaged d$I$/d$V$ for the epi-ready SrTiO$_3$ shows the highest density of states and largest standard deviations because of the high defect density and a wide distribution of peak bias. The averaged spectrum for the 10-u.c.-thick SrTiO$_3$ film is intermediate between that for the epi-ready sample and the 50-u.c.-thick film because of the reduced number of defects and a narrower peak distribution than the epi-ready sample. In the 50-u.c.-thick film, in contrast, the spectral weight is significantly reduced and the standard deviations are considerably smaller than for the other samples, implying an electronically homogeneous surface. Thus, we conclude that defects associated with oxygen vacancies were incorporated into the SrTiO$_3$ substrate and the ultrathin layer of SrTiO$_3$, but by growing a thick film, those defects could be effectively precluded from the film surface.

Our finding of the thickness dependence of the defect density in SrTiO$_3$ strongly indicates the significant role of defects that reside near the interfaces of oxide heterostructures. The non-uniform stoichiometry along the growth direction should be studied more carefully using various techniques to fully understand the many unique properties at oxide heterointerfaces. In addition, the initial growth process of oxide thin films needs to be addressed in more detail. Low-temperature STM/STS combined with PLD offers a unique means of studying the atomic-scale electronic properties of oxide thin films, and furthermore, can be employed for functional oxide systems such as colossal magnetoresistive manganites and high-$T_c$ superconductors. We believe that such studies will lead to a new approach for engineering novel functional complex oxides.

**Conclusions**

Clean surfaces of a SrTiO$_3$ epi-ready substrate and homoepitaxial thin films grown in layer-by-layer mode were examined using low-temperature STM/STS-PLD for the first time. Annealing a buffered-HF-etched single-crystal SrTiO$_3$(001) substrate at 1000˚C in an oxygen partial pressure of $1\times10^{-6}$ Torr, which is a well-known method to provide a smooth TiO$_2$-terminated SrTiO$_3$ surface, resulted in atomically disordered structures, despite the formation of equidistant straight step edges. In contrast, homoepitaxial SrTiO$_3$ thin film surface with thicknesses of 10-u.c. and 50-u.c. exhibited a (2×2) surface reconstruction. The STS measurements revealed the presence of defects on the epi-ready and 10-u.c.-thick film surfaces, and that the defect density drastically decreased in thicker films, thus resulting in thickness-dependent inhomogeneous electronic structures which originate from surface defects. These results indicate that the stoichiometry of the film is not uniform along the growth direction.

**Acknowledgement**

This study was supported by the World Premier Research Institute Initiative, promoted by the Ministry of Education, Culture, Sports, Science, and Technology (MEXT), Japan, for the Advanced Institute for Materials Research, Tohoku University, Japan.

**Figure captions**

**Figure 1.** (a) Wide-area STM image (250 nm × 250 nm) and (b) height profiles measured normal to the step edge. Images were obtained at 78 K under a sample bias $V_s$ of +1.5 V and tunneling current $I_t$ of 30 pA. The inset in (a) shows the corresponding RHEED pattern. (c) Magnified STM image of a typical $TiO_2$-terminated $SrTiO_3$(001) surface (50 nm × 50 nm) taken at $V_s$ = +2.8 V and $I_t$ = 30 pA.

**Figure 2.** RHEED intensity oscillations of the specular reflected beam monitored during the growth of $SrTiO_3$ on the $TiO_2$-terminated $SrTiO_3$(001) substrate surface. (a) layer-by-layer 10-u.c.- and (b) 50-u.c.-thick $SrTiO_3$ growth at a $T_g$ of 700˚C. (c), (d) Wide-area (150 nm × 150 nm) and (e), (f) magnified STM images (20 nm × 20 nm) of 10-u.c.- and 50-u.c.-thick $SrTiO_3$ films, respectively. (e) $V_s$ = +2.0 V, $I_t$ = 20 pA, (f) $V_s$ = +3.0 V, $I_t$ = 30 pA. The inset of (e) is a fast Fourier transformed image of the STM image in (e).

**Figure 3.** (a) STM image (20 nm × 20 nm) for 10-u.c.-thick $SrTiO_3$ at $V_s$ = +3.5 V taken simultaneously in STS measurement ($I_t$ = 30 pA). (b), (c) Differential conductance $dI/dV$ maps in the same field of view at $V_s$ = +2.5 V, 2.9 V, respectively. (d) Spatially averaged and normalized $dI/dV$ spectra taken on defect sites in the $dI/dV$ maps, where the colors correspond to those in (a)-(c). (e) Areal defect density plotted as a function of $V_s$ in epi-ready (upper) and 10-u.c.-thick $SrTiO_3$ (bottom).

**Figure 4.** Comparison of $dI/dV$ maps ($V_s$ = +2.7 V) between (a) 50-u.c.- and (b) 10-u.c.-thick $SrTiO_3$ films. (c) Averaged $dI/dV$ spectra taken in the empty state for epi-ready, 10-u.c.- and 50-u.c.-thick $SrTiO_3$ samples.

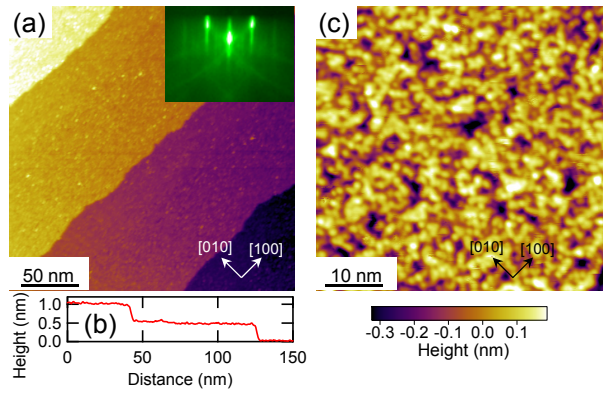

Figure 1 Ohsawa et al

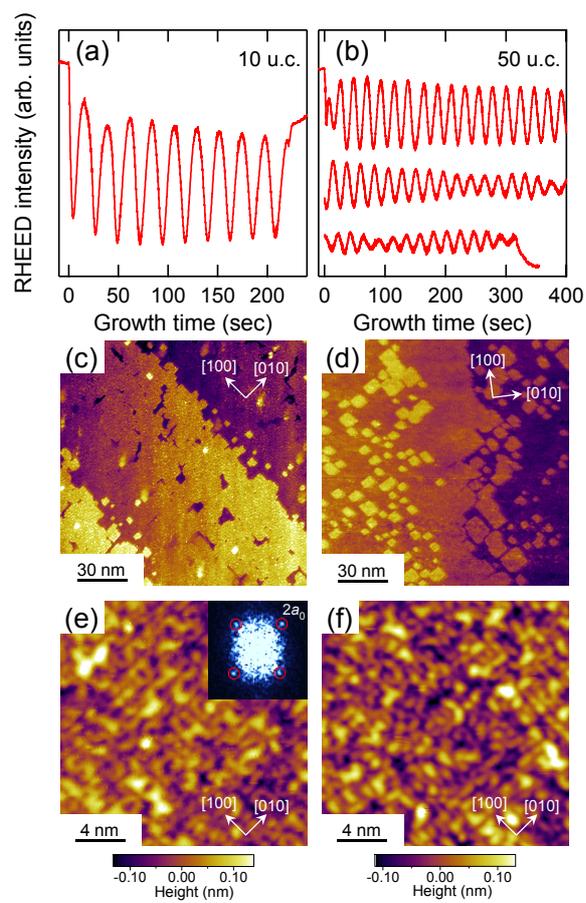

Figure 2 Ohsawa et al.

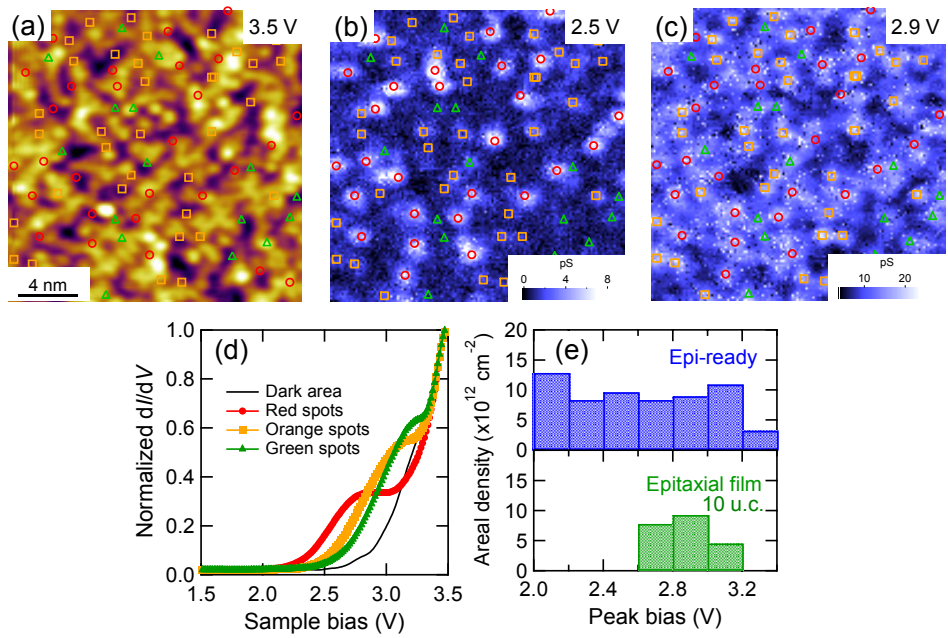

Figure 3 Ohsawa et al.

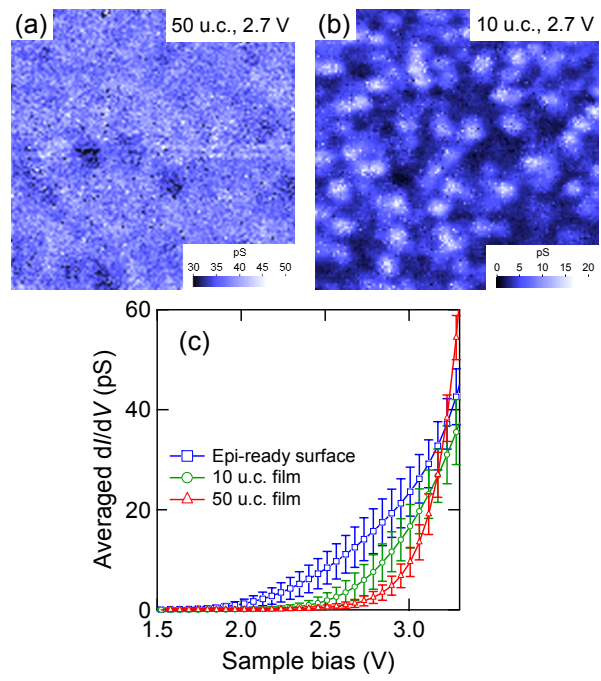

Figure 4 Ohsawa et al.